\documentclass[aps,prb,floats,twocolumn]{revtex4}

\usepackage{amsmath}
\usepackage{epsfig}
\begin{document}

\title{Stabilization of d-Band Ferromagnetism by Hybridization with Uncorrelated Bands} 
\author{S. Schwieger and W. Nolting}
\begin{abstract}
We investigate the influence of s-d or p-d hybridization to
d-band ferromagnetism to estimate the importance of hybridization for the
magnetic properties of transition metals. To focus our attention to the
interplay between hybridization and correlation we investigate a simple
model system consisting of two non-degenerated hybridized bands, one strongly
correlated, the other one quasi-free. To solve this extended Hubbard
model, we apply simple approximations,
namely SDA and MAA, that, concerning ferromagnetism in the single-band
model, are known to give qualitatively satisfactory
results. This approach allows us to discuss the underlying mechanism, by
which d-band ferromagnetism is influenced by the hybridization on the
basis of analytical expressions. The latter clearly display the order and the
functional dependencies of the important effects. It is found, that
spin-dependent inter-band particle fluctuations cause a spin-dependent band
shift and a spin-dependent
band broadening of the Hubbard bands. The shift stabilizes, the
broadening tends to destabilize ferromagnetism. Stabilization requires
relatively high band distances and small hybridization matrix elements. Super-exchange and RKKY coupling are of minor importance. 
\end{abstract}
\maketitle
\section{Introduction}
\label{intro}
The issue of magnetism in band-ferromagnets like Fe, Co and Ni is far
from being settled. Magnetism in this materials is due to correlations within 
itinerant electron bands. The simplest model that comprises this aspects
is the single-band Hubbard model. Although it was introduced to gain a
first qualitative understanding of band-ferromagnetism\cite{GUT63}$\,$\cite{HUB63}$\,$\cite{KAN63} it took almost
30 years to answer the question whether it is a generic model for
ferromagnetism at all. About ten years ago a Dynamical Mean Field
Theory (DMFT)\cite{MEV89}$\,$ \cite{VOL89}$\,$ \cite{JAR92}$\,$ \cite{GKKR96} was developed, which allows a consistent (mean field)
description of the whole parameter range of the single-band Hubbard
model. DMFT-based calculations confirmed the existence of ferromagnetism
for a wide parameter range\cite{JP93}$\,$\cite{VBHK97}$\,$\cite{ULM98}. Today there is a general consensus that the single-band
Hubbard model exhibits ferromagnetism.\\
There is also consensus, however, that this model oversimplifies the
situation in band-ferromagnets, for instance by restricting the
correlations to the on-site elements. But an even more drastic simplification
is the restriction to a single non-degenerated electron band. The
fivefold degeneracy of the d-electrons certainly influences the magnetic
properties of the system. Consequently, a lot of effort is being done by
transferring certain treatments, once developed for the single-band
model, to multi-band models. Let us mention Gutzwiller approximation
\cite{BWG98} or
various treatments within the DMFT-frame\cite{LK98}$\,$\cite{MZPK99}.\\
Besides the degeneracy of the d-electrons, the single-band model also
neglects weakly correlated $s$- and $p$-bands, although they are located around the Fermi energy in $3d$
transition-metals. The interplay between correlated and uncorrelated
electrons is known to give rise to a variety of phenomena such as the
Kondo effect or heavy fermions\cite{HEW93} and is the central point of widely used
models like the Anderson model. In the case of the periodic Anderson
model (PAM), correlations in combination with the hybridization to an
uncorrelated band can cause
ferromagnetism\cite{MW93}$\,$\cite{TJF97}$\,$\cite{MN00}
as shown rigorously for the one-dimensional case at quarter filling\cite{YAS93}.  This indicates that
uncorrelated bands may influence the magnetic phase diagram of the
Hubbard model, too, and this is most likely if the band distance is smaller than the on-site
Coulomb energy (charge transfer regime\cite{CK94}). Recent experiments indeed
seem to indicate that ferromagnetism can be stabilized if additional
p-orbitals are doped into a ${\rm RECo_2}$-system(${\rm RE=Ho,Er}$)\cite{CHSAAKD97}. The aim of this paper is
to decide, whether the neglect of $s$- and $p$-bands is justified when modeling band-ferromagnets like Fe, Co, or Ni. The influence of the
hybridization of $d$-electrons with these orbitals shall be investigated systematically.\\\\
The paper is organized as follows. In the next section a suitable
Hamiltonian is formulated and we try to give a qualitative overview of
the interplay between the two different kinds of electrons. In section \ref{theory} we will
apply certain approximations to the Hamiltonian. Thereby we will try to get as
much insight as possible into the mechanisms, by which the $d$-band
magnetism is altered. While the above mentioned DMFT-based
treatments give certainly reliable values for magnetic properties, it
is challenging to give a direct physical meaning to auxiliary
quantities used in this theory (e.g. the energy- and spin-dependent hybridization function). For these
reasons we will formulate the much simpler Hubbard-I decoupling ({\it Hu-I})\cite{HUB63}, the spectral
density approach ({\it SDA})\cite{NOL72}, and the modified alloy analogy ({\it MAA})\cite{HEN96} for
the described multi-band model. These theories are conceptually restricted to high energy excitations in
the strong coupling regime. This is, however, the interesting case
concerning band-ferromagnetism.  For the single-band model in the limit
of infinite spatial dimensions the theories are thoroughly tested
against numerical
exact results available in this limit\cite{PHW98}. It is found that the {\it SDA} as well as the {\it MAA}
systematically overestimate magnetic quantities such as the Curie
temperature but turn out to give a qualitative satisfying description of
band-ferromagnetism\cite{PHW98}$\,$\cite{MN00}. For our purpose the main advantage of
these theories is the possibility for analytical estimations. \\
In section \ref{results} the main results concerning the p-band
influence on ferromagnetism are
shown. Within the {\it SDA} we will derive analytical expressions for
the quasiparticle band structure in the strong coupling limit. This
allows a vivid physical interpretation of the mechanism by which the
properties of the correlated subsystem are influenced by uncorrelated
bands. We will see that the main impact is due to spin dependent
inter-band fluctuations, which may enhance or reduce the spin asymmetry
of the interacting density of states. Finally we discuss alternative
mechanisms that involve the new states, like superexchange and RKKY-coupling.

\section{General considerations}
\label{general}
We want to study the influence of weakly correlated bands to 
$d$-band ferromagnetism within the following extension of the
single-band Hubbard model:
\begin{eqnarray}
H&=&\sum_{ij\sigma}(T_{ij}^d-\mu)d_{i\sigma}^+d_{j\sigma}+\frac{U}{2}\sum_{i\sigma}n_{i\sigma}^dn_{i-\sigma}^d\nonumber\\
&{}&+\sum_{ij\sigma}(T^p_{ij}-\mu)p_{i\sigma}^+p_{j\sigma}+V\sum_{i\sigma}(d_{i\sigma}^+p_{i\sigma}+p_{i\sigma}^+d_{i\sigma})\quad
.
\label{Ham}
\end{eqnarray}
This Hamiltonian is similar to those used e.g. in reference \cite{CK94} and reduces
to the periodic Anderson model (PAM) in the limit $T_{ij}^d\rightarrow 0$ for
$i\neq j$. 
The weakly correlated electrons are described by a quasi-free
"$p$-band", with the hopping integrals $T_{ij}^p$, while the single-band
Hubbard model describes the $d$-system. $T_{ij}^d$ are the hopping
integrals within the $d$-band and $U$ is the local Coulomb interaction.
The bands are coupled by a hybridization $V$.    
The hopping integrals are the Fourier transformed Bloch energies 
and $\mu$ denotes the chemical potential.
The free band structure $\epsilon_k^{p;d}$ shall be the result of a
tight-binding approximation. The relative position of the bands is
characterized by two parameters: The difference of the free centers of
gravity $\Delta T_0$ and the ratio of the free band
widths $\alpha$:
\begin{equation}
\Delta T_0=T_{0}^p-T_{0}^d\quad\quad \alpha=\frac{{\cal
    W}_0^p}{{\cal W}_0^d}\quad.
\label{definition}
\end{equation}
$T_0^{p,d}=T_{ii}^{p,d}$ are the centers of gravity of the free bands.
To achieve a realistic description of transition metals we choose
$\alpha>1$ and $\Delta T_0>0$. As a consequence of the tight binding
approximation the dispersions are connected via
\begin{equation}
\epsilon_k^p=T_0^p+\alpha\cdot(\epsilon_k^d-T_0^d)\quad .
\label{alpha}
\end{equation}
Let us now discuss the possible influences of the $p$-band on the
$d$-system within this model.\\
First of all, there is a rather trivial particle number
effect\cite{BJ79}.
Magnetism depends sensitively on the $d$-particle density.
 If now the
new band is added while the total particle number in the system stays
fixed, the electron density within the correlated subsystem is
changed. The same holds if the parameters $V$ or $\Delta T_0$ are tuned.
We do not want to address these effects here. Note, that our intention is
not to describe effects resulting from an experimental tuning of the
hybridization strength, e.g. by applying pressure. Rather we want to
decide if the neglect of the $s,p$-$d$ hybridization is a good
approximation for many body model calculations. In this context it is
assumed that even when the $s$- and $p$-electrons are neglected the
correct $d$-particle number per site is used. This generally non-integer
number is already the result of the hybridization to other bands. Thus
we will regard this case (where the change of the $d$-particle
number due to the hybridization is already considered) and the case of an explicitely treated
hybridization (where additionally all other effects resulting from the
two-band situation are taken into account). To compare these cases properly we have to fix the
$d$-particle density in our calculations. 
\\
What further effects can be expected? Naively, one would believe that
an uncorrelated and therefore a priori "non-magnetic" $p$-band would
destabilize ferromagnetism by "reducing the average
correlation". This reasoning, however, is too simple. Particle fluctuations between
the bands will influence the propagation of electrons within the
$d$-band and thus the $d$-projected density of states. It is known that
ferromagnetism depends sensitively on the the shape of the density of
states\cite{WBS98}. This effect will be most important if the
fluctuation rate is spin-dependent. This would cause different
alterations of the spin-up and spin-down density of states and directly
influence its spin asymmetry.
\\
Let us look at this mechanism in the trivial limiting case of
uncorrelated bands $U\rightarrow 0$. For small hybridizations the
excitation energies are:
\begin{eqnarray}
E_{k1}(V)&=&\epsilon_k^d-\frac{V^2}{|\epsilon_k^p-\epsilon_k^d|}\nonumber\\
E_{k2}(V)&=&\epsilon_k^p+\frac{V^2}{|\epsilon_k^p-\epsilon_k^d|}
\label{dispersionen}
\end{eqnarray}
For the lower band $E_{k1}(V)$ this causes a band asymmetry, a band shift to lower
energies, and a band broadening in the quasiparticle density of states.
For non-overlapping bands, i.e.
$\Delta T_0>max(\epsilon_k^{p}-T_0^{p})$, we insert
(\ref{alpha}) into (\ref{dispersionen}) and expand
$E_{k1}(V)$ in powers of
$\frac{(\alpha-1)(\epsilon_k^d-T_0^d)}{\Delta
  T_0}$. Equation (\ref{dispersionen}) becomes 
\begin{eqnarray}
E_{k1}(V)=T_0^d+\Delta T_V^d+\left( \epsilon_k^d-T_0^d\right)\cdot x_V^d
\label{dispersionen2}
\end{eqnarray}
with the band shift
\begin{equation}
\Delta T_V^d=-\frac{V^2}{\Delta T_0}
\label{shift}
\end{equation}
and the band broadening factor
\begin{equation}
x_V^d= 1+\frac{V^2}{\Delta T_0^2}(\alpha-1)\quad.
\label{broad.}
\end{equation}
The broadening as well as the shift are also present if the
$d$-electrons are correlated as can be seen by studying a two-site cluster out of (\ref{Ham})
with the inter-site hoppings $t^d$ and $t^p=\alpha\cdot t^d$. For small $V$ one can
perform a canonical transformation that decouples the $p$- and $d$-band in
first order in $V$. The calculation is lengthy but straightforward. For $U\rightarrow\infty$, $\Delta T_0>
t^{(p;d)}$, and $T_0^d<\mu<T_0^p$ the
$d$-electrons are well approximated by a two-site Hubbard Hamiltonian
with a renormalized center of gravity $\hat{T}_0^d(V)$,
renormalized hopping integrals $\hat{t}^d(V)$, and a renormalized interaction $\hat{U}(V)$.
We find the parameters
\begin{eqnarray}
\hat{T}_0^d(V)&=&T_0^d-\frac{V^2}{\Delta T_0}\nonumber\\
\hat{t}^d(V)&=&t^d(1+\frac{V^2}{\Delta T_0^2}(\alpha-1))\nonumber\\
\hat{U}(V)&=&U+\frac{V^2}{\Delta T_0}\quad.
\label{cluster}
\end{eqnarray}
The broadening as well as the shift is clearly recognized in
(\ref{cluster}). 
Our preceding, qualitative considerations indicate that
inter-band particle fluctuations indeed modify the
d-projected density of states. These modifications are expected to influence also
the magnetic properties. Up to now we only investigated
spin-symmetric limiting cases allowing only a spin symmetric fluctuation
rate. Regarding ferromagnetism it will be most
important whether one of the effects becomes spin-dependent in the full system.

\section{Theory} 
\label{theory}
The magnetic properties of (\ref{Ham}) can be studied using retarded single electron Green functions
\begin{eqnarray*}
G_{k\sigma}^{dd}&=&\langle\langle
d_{k\sigma};d_{k\sigma}^+\rangle\rangle\quad
G_{k\sigma}^{pp}=\langle\langle
p_{k\sigma};p_{k\sigma}^+\rangle\rangle\\
G_{k\sigma}^{dp}&=&G_{k\sigma}^{pd}=\langle\langle
d_{k\sigma};p_{k\sigma}^+\rangle\rangle =\langle\langle
p_{k\sigma};d_{k\sigma}^+\rangle\rangle\quad ,
\end{eqnarray*}
which fulfill the following equations of motion
\footnote{natural units are used throughout this paper, hence $\hbar=1$}:
\begin{eqnarray}
EG_{k\sigma}^{dd}&=&1+(\epsilon_k^d-\mu)G_{k\sigma}^{dd}+\Sigma_{k\sigma}G_{k\sigma}^{dd}+VG_{k\sigma}^{pd}\nonumber\\
EG_{k\sigma}^{pd}&=&(\epsilon_k^p-\mu)
G_{k\sigma}^{pd}+VG_{k\sigma}^{dd}\nonumber\\
EG_{k\sigma}^{pp}&=&1+({\epsilon}_k^p-\mu)G_{k\sigma}^{pp}+VG_{k\sigma}^{dp}\quad
.
\label{EOM}
\end{eqnarray}
The self-energy $\Sigma_{k\sigma}$ is introduced as usual via
\begin{equation}
\Sigma_{k\sigma}G_{k\sigma}^{dd}=\langle\langle
[d_{k\sigma},\frac{U}{2}\sum\limits_{i\sigma}n_{i\sigma}^d
n_{i-\sigma}^d]_-;d_{k\sigma}^+\rangle\rangle\quad ,
\label{def.sigma}
\end{equation}
where $[\ldots ,\ldots]_-$ denotes the commutator.
Solving (\ref{EOM}) gives all Green functions: 
\begin{equation}
\left(\begin{array}{cc} G^{dd}_{k\sigma} &
    G^{dp}_{k\sigma}\\
  {}&{}\\
      G^{pd}_{k\sigma} &
      G^{pp}_{k\sigma}\end{array}\right)^{-1}=\left(\begin{array}{cc}
      E-\epsilon_k^{\prime p} & -V\\
{}&{}\\
       -V & E-\epsilon_k^{\prime d}-\Sigma_{k\sigma}\end{array}\right)\!,
\label{matrix}
\end{equation}
where ${\epsilon}_k^\prime$ is used as an abbreviation for $\epsilon_k-\mu$.\\
In the ferro- and paramagnetic phase we can calculate the spin-dependent average occupation
numbers $n_\sigma^d=\langle n_{i\sigma}^d\rangle$ and
$n_\sigma^{p}=\langle n_{i\sigma}^p\rangle$ using the Green functions
(\ref{matrix}):
\begin{eqnarray}
n_{\sigma}^{(p;d)}=-\frac{Im}{\pi}\int_{-\infty}^\infty dE\,
f_-(E)G_{ii\sigma}^{(pp;dd)}(E-\mu)\quad.
\label{nvonro}
\end{eqnarray}
$f_-(E)$ is the Fermi function and $G_{ii\sigma}^{(pp;dd)}$ are the local
Green functions.\\
Obviously we can calculate the phase boundary between para- and
ferromagnetism as soon as we have
found an (approximate) expression for the self-energy.\\  
To this aim we will formulate the {\it Hu-I}, {\it SDA}, and {\it MAA} for
the two-band problem (\ref{Ham}). By comparing the influence of the
hybridization within different approximations we can minimize the risk of
an artificial $p$-band influence. The two "simple" approximations {\it
  Hu-I} and {\it SDA} can give an excellent insight into the working
mechanisms. Due to their explicit structure of the self energy,
one can perform some demonstrative analytical estimations. The {\it SDA} gives
qualitatively convincing results concerning ferromagnetism. This is due to
the fact that it reproduces the correct values for the centers of
gravity and weights of the Hubbard bands in the strong coupling limit
$U\rightarrow\infty$. Compared to {\it Hu-I}, an additional correlation
function is considered that describes the itineracy of electrons of
opposite spin direction and allows for a spin-dependent band shift.
The {\it MAA} is a first systematic improvement of the {\it SDA}, since it
allows quasiparticle damping, which is completely neglected within the
{\it SDA}. By comparing {\it MAA}- and {\it SDA}-results one can see,
if the mechanisms derived within the {\it  SDA} are also present in a
more complex theory. 
\\\\
\textbf{Hubbard-I decoupling}: 
Let us start with the Hubbard-I approximation. Straightforward
decoupling of the real space equations of motion in (\ref{matrix}) yields the {\it Hu-I}
self-energy
\begin{equation}
{}^{{\it
    Hu-I}}\Sigma_\sigma=Un_{-\sigma}^d\frac{E-T_0^d+\mu}{E-T_0^d+\mu-U(1-n_{-\sigma}^d)}\quad ,
\label{hu-i}
\end{equation}
which is formally identical to the single-band case. The self-energy is
$V$-dependent via $n_{-\sigma}^d$, which is calculated using (\ref{matrix}) and
(\ref{nvonro}). Equation (\ref{hu-i}) gives three excitation energies in
every point of the Brillouin-zone, corresponding to the three-peak
structure of the spectral density in the atomic limit $V\rightarrow 0; T_{ij}^{i\neq
  j}\rightarrow 0$. Finite values of the hopping and hybridization change the
positions and weights of the $\delta$-peaks and lead to a mixing of $p-$ and
$d-$ spectral density.\\\\
\textbf{SDA}
For the single-band model, the {\it SDA} is the simplest theory that yields the correct strong
coupling and high energy behaviour, which seems to be decisive for the
existence of ferromagnetism. The general structure of the spectral density and the
self-energy is the same as in {\it Hu-I}. The energy-positions and weights of the $\delta$-peaks in the spectral density are obtained by
fitting it to the first four spectral moments:
\begin{eqnarray}
{}^{dd}M_{k\sigma}^{(m)}&=&\int_{-\infty}^\infty\, dE\, E^m
S_{k\sigma}^{dd}(E)\nonumber\\
&=&\langle
[\underbrace{[[[d_{k\sigma},H]_-H]_-\ldots H]_-}_{m-fold},d_{k\sigma}^+]_+\rangle\quad .
\label{momente}
\end{eqnarray}
$[\ldots ,\ldots]_+$ is the anti-commutator.
For the two-band model we will apply this concept directly to the
self-energy rather than to the spectral density. Therefore we choose the
same structure as in (\ref{hu-i}) for a self-energy ansatz:
\begin{equation*}
{}^{\it
  SDA}\Sigma_{k\sigma}\stackrel{!}{=}\gamma_1\frac{E-\gamma_2}{E-\gamma_3}\quad .
\label{ansatz}
\end{equation*}
The parameters $\gamma_i$ shall now be fitted to the spectral moments.
To this end we expand the Green function and the
self-energy with respect to powers of $\frac{1}{E}$:
\begin{eqnarray}
G_{k\sigma}^{dd}=\sum_{n=0}^\infty\frac{{}^{dd}M_{k\sigma}^{(n)}}{E^{n+1}};\quad
\Sigma_{k\sigma}=\sum_{n=0}^\infty \frac{C_{k\sigma}^{(n)}}{E^n}\quad,
\label{hochenergie}
\end{eqnarray}
The high-energy coefficients of the Green function are the spectral
moments (\ref{momente}). This can easily be seen by expanding the
spectral representation of the Green function
\begin{equation*}
G_{k\sigma}^{dd}(E)=\int_{-\infty}^\infty
dE^\prime\frac{S_{k\sigma}^{dd}(E^\prime)}{E-E^\prime +i0^+}
\end{equation*}
with respect to $\frac{1}{E}$ and comparing the resulting expressions
with the definition of the moments (\ref{momente}).
The self-energy coefficients
$C_{k\sigma}^{(n)}$ are obtained as functions of the moments
${}^{dd}M_{k\sigma}^{(0)}\ldots {}^{dd}M_{k\sigma}^{(n+1)}$ by inserting
the expansions (\ref{hochenergie}) into (\ref{matrix}) (or equivalently
into the
Dyson equation) and by comparing the
coefficients of the $\frac{1}{E^n}$ terms. With the r.h.s. of
(\ref{momente}), we find the first four correlated spectral moments:
\begin{eqnarray}
{}^{dd}M_{k\sigma}^{(0)}&=&1\nonumber\\
{}^{dd}M_{k\sigma}^{(1)}&=&\epsilon_k^{\prime d}+Un_{-\sigma}^d\nonumber\\
{}^{dd}M_{k\sigma}^{(2)}&=&(\epsilon_k^{\prime
  d})^2+2Un_{-\sigma}^d\epsilon_k^{\prime d}+U^2
n_{-\sigma}^d+V^2\nonumber\\
{}^{dd}M_{k\sigma}^{(3)}&=&(\epsilon_k^{\prime
  d})^3+3Un_{-\sigma}^d(\epsilon_k^{\prime d})^2+U^2(2n_{-\sigma}+(n_{-\sigma}^d)^2)\nonumber\\&{}&+U^2n_{-\sigma}^d(1-n_{-\sigma}^d)(B_{k-\sigma}^{2-band}+T_0^\prime)\nonumber\\ 
&{}&+U^3n_{-\sigma}^d+V^2(2\epsilon_k^{\prime d}+\epsilon_k^{\prime p}+2Un_{-\sigma}^d)\quad.
\label{zweib-moments}
\end{eqnarray}
The self-energy coefficients read:
\begin{eqnarray*}
C^{(0)}_\sigma&=&Un_{-\sigma}^d\nonumber\\
C^{(1)}_\sigma&=&U^2n_{-\sigma}^d(1-n_{-\sigma}^d)\nonumber\\
C^{(2)}_{k\sigma}&=&U^2n_{-\sigma}^d(1-n_{-\sigma}^d)(B_{k-\sigma}^{2-band}+T_0^\prime+U(1-n_{-\sigma}^d))\quad.
\label{koeffizienten}
\end{eqnarray*}
$B_{k\sigma}^{2-band}=B_\sigma^{2b}+F_{k\sigma}^{2b}$ is a higher correlation function with the local
part $B_\sigma^{2b}$ and a $k-$dependent part
$F_{k\sigma}^{2b}$. For the single-band model, the influence of  both
terms is discussed in detail in reference \cite{HEN97A}. It turns out that the
most important term is the local $B_\sigma$, which leads to a
spin-dependent band shift in the ferromagnetic phase. With regard to
ferromagnetism the non-local part $F_{k\sigma}$ seems to be of minor
importance. Therefore we will neglect it in the following. From (\ref{momente}) we find for the local part $B_\sigma^{2b}$:
\begin{eqnarray*}
B_{\sigma}^{2b}&=&\frac{1}{n_\sigma^d(1-n_\sigma^d)}\Bigg(\frac{1}{N}\sum_{ij}^{i\neq
  j}T_{ij}^d\langle d_{i\sigma}^+ d_{j\sigma}(2n_{i-\sigma}^d-1)\rangle\\
&{}&+\frac{2V}{UN}\sum_{ij}(T^p_{ij}-T_{ij}^d)\langle
d_{i\sigma}^+p_{j\sigma}\rangle\\
&{}&-V\langle d_{i\sigma}^+
  p_{i\sigma}\rangle+\frac{2V^2}{U}(n_\sigma^d-n_\sigma^p)\Bigg)\quad.\nonumber 
\end{eqnarray*}
Although $B_{\sigma}^{2b}$ contains expectation
values of the uncorrelated band and higher correlation functions, it can
be expressed as a functional of the correlated single-electron
Green function only:
\begin{eqnarray}
B_\sigma^{2b}&=\frac{-Im}{n_\sigma(1-n_\sigma)\pi}\int_{-\infty}^\infty dE\, 
f_-(E)\left(\frac{2
    \Sigma_\sigma(E)}{U}-1\right)\cdot\nonumber\\
&{}\!\!\cdot\left((E-\Sigma_\sigma(E)-T_0)G_{ii\sigma}^{dd}(E-\mu)-1\right).
\label{Be}
\end{eqnarray}
The correlation-function $B_\sigma^{2b}$ and the
self-energy coefficients $C_\sigma^{(0;1;2)}$ turn out to be the same
functionals of the correlated Green function $G_{ii\sigma}^{dd}$ as in the single-band model. While determining the self-energy
coefficients, the whole $V$-dependence in the
moments ${}^{dd}M_{k\sigma}^{(0)}\ldots {}^{dd}M_{k\sigma}^{(3)}$
(\ref{zweib-moments}) is cancelled by the explicit
$V$-dependence of the correlated Green function (\ref{matrix}). 
Thus, like in the {\it Hu-I} approximation,  the {\it SDA} self-energy is
formally identical to the single-band case:
\begin{equation*}
{}^{\it
  SDA}\Sigma_\sigma=Un_{-\sigma}^d\frac{E-T_0+\mu-B_{-\sigma}^{2b}}{E-T_0+\mu-B_{-\sigma}^{2b}-U(1-n_{-\sigma}^d)}\quad .
\end{equation*}
The $V-$dependence comes again into play by the expectation values
$n_{-\sigma}$ and $B_{-\sigma}^{2b}$ being evaluated via
(\ref{matrix}), (\ref{nvonro}), and (\ref{Be}).\\\\
\textbf{MAA} Besides the restriction to strong interaction strengths a drawback of {\it SDA} and {\it Hu-I} is the exclusion of scattering
processes, which lead to quasiparticle damping. The correlated $d$-system
is described by two quasiparticles with infinite lifetime corresponding
to singly or doubly occupied sites. One possibility to include quasiparticle damping is the
description of the system by a fictitious alloy (alloy analogy), which is a standard
method in many body physics\cite{VKE68}. With this approach one can
account for electron scattering at the potentials formed by the
distribution of the electrons of opposite spin direction. Main excitation energies of the many body
system are modelled by atomic energy levels of a fictitious
alloy. Correlation effects are then described by the properties of this
alloy and its self-energy is identified with the
self-energy of the many body problem. Since the self-energy
(\ref{def.sigma}) exclusively characterizes correlated electrons, we will only
describe the correlated subsystem by a fictitious alloy. In the strong coupling limit we have two main excitation
energies within the correlated subsystem. Consequently we will describe it
by a two-component alloy. The resulting effective alloy problem can be
solved by the Coherent Potential Approximation (CPA), which yields the CPA self-energy
\begin{equation}
0=\sum_{p\sigma}^{p=1,2}
x_{p\sigma}\frac{E_{p\sigma}-T_0^d-\Sigma_\sigma(E)}{1-G_{ii\sigma}^{dd}(E)(E_{p\sigma}-\Sigma_\sigma(E)-T_0^d)}\quad.
\label{CPA}
\end{equation}
$E_{p\sigma}$ and $x_{p\sigma}$ are the atomic energy-levels and
concentrations of the alloy components. The  CPA, being a single-site
theory, gives a local self-energy $\Sigma_\sigma$.
After rearranging the terms and setting
\begin{eqnarray*}
\gamma_1&=&x_{1\sigma}(E_{1\sigma}-T_0^d)+x_{2\sigma}(E_{2\sigma}-T_0^d)\\
\gamma_2&=&\frac{(E_{1\sigma}-T_0^d)(E_{2\sigma}-T_0^d)}{\gamma_1}\\
\gamma_3&=&x_{1\sigma}(E_{2\sigma}-T_0^d)+x_{2\sigma}(E_{1\sigma}-T_0^d)\quad,
\end{eqnarray*}
equation (\ref{CPA}) becomes
\begin{eqnarray}
\Sigma_\sigma(E)=\gamma_1\frac{1+G_{ii\sigma}^{dd}(E)(\Sigma_\sigma(E)-\gamma_2)}{1+G_{ii\sigma}^{dd}(E)(\Sigma_\sigma(E)-\gamma_3)}\quad.
\label{CPA-3}
\end{eqnarray}
Here $x_{1\sigma}+x_{2\sigma}=1$ is already used. To complete the
theory, we now have to adjust the parameters $\gamma_1,\gamma_2$, and $\gamma_3$. Similar to the {\it SDA} these parameters can
be fitted to the on-site
spectral moments $M_{ii\sigma}^{(m)}$ and on-site self-energy
coefficients $C_{ii\sigma}^{(m)}$. The two latters are defined analogously to
$M_{k\sigma}$ and $C_{k\sigma}$ (\ref{momente}). To this purpose one has to expand
the local Green function $G_{ii\sigma}^{dd}$ and the local self-energy
$\Sigma_\sigma$ in powers of $\frac{1}{E}$ analogously to
(\ref{hochenergie}). Then one inserts these expansions into (\ref{CPA-3})
and compares the coefficients of the $\frac{1}{E^n}$-terms up to $n=2$.\footnote{best to be done in the form $\Sigma_\sigma-\gamma_1+G_{ii\sigma}[\Sigma_\sigma(\Sigma_\sigma-\gamma_3)-\gamma_1(\Sigma_\sigma-\gamma_2)]=0$}
Using the abbreviation ${}^{MAA}\Sigma_\sigma\rightarrow
\Sigma_\sigma$, we finally
find for the {\it MAA} self-energy:
\begin{equation*}
\Sigma_\sigma=Un_{-\sigma}^d\frac{(G_{ii\sigma}^{dd})^{-1}+\Sigma_\sigma-B_{-\sigma}^{2b}}{(G_{ii\sigma}^{dd})^{-1}+\Sigma_\sigma-B_{-\sigma}^{2b}-U(1-n_{-\sigma}^d)}
\end{equation*}
This is again, as in {\it Hu-I} and {\it SDA}, formally identical to
the single-band expression, i.e. the self-energy is the same functional of
the correlated Green function as in the single-band case. The self-energy is $V$-dependent via
$G_{ii\sigma}^{dd}$ and the
expectation-values $n_\sigma^d$ and $B_\sigma^{2b}$.\\
The {\it MAA} self-energy is still consistent with the high-energy limit
and additionally allows for quasiparticle damping, thus being a systematic
improvement of the {\it SDA}\footnote{However, as in the {\it SDA}, the
  low energy behaviour is described poorly. For instant the imaginary part of the self
  energy does not vanish at the Fermi energy.}.
\begin{figure}
\begin{center}
      \epsfig{file=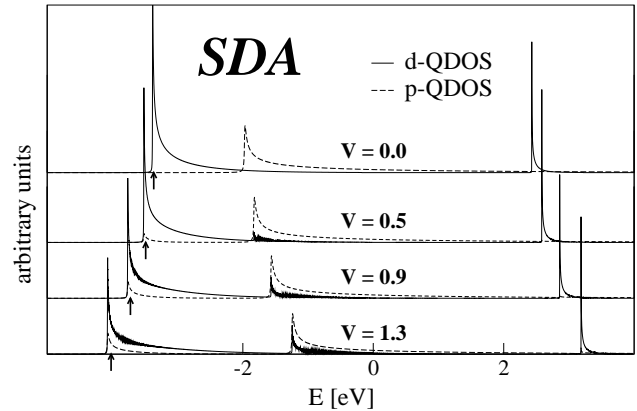,angle=270,width=1.0\linewidth}
\caption{Quasiparticle density of states (QDOS) as a function of the energy E at different V. The
  system is forced to be paramagnetic. Parameters: $U=5.0, n=0.25,
  \alpha=4.0, \Delta T_0=3.0, T=0K, fcc(\infty)$-lattice, see eqn.(\ref{DOS}). The arrows show the position of the
  Fermi-energy. With rising $V$ the distance between the sub-bands
  increases quadratically.}
\label{para}
\end{center}
\end{figure}

\section{Results and Discussion}
\label{results}
\begin{figure}
\begin{center}
      \epsfig{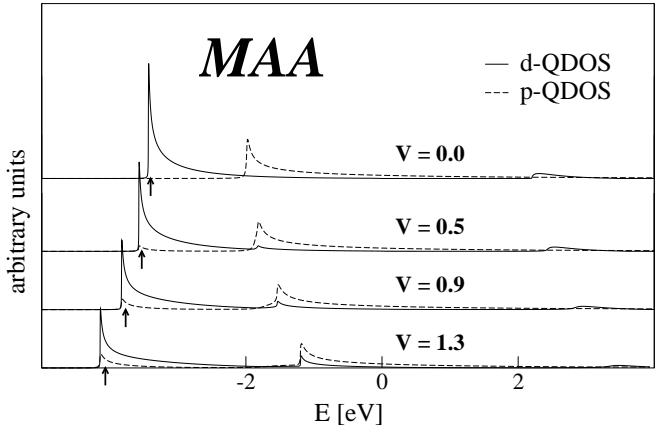}
\caption{Same as in Fig. \ref{para}, but calculated with {\it MAA}. The
  peaks are broader than in Fig. \ref{para} due to quasiparticle damping.
  This is most pronounced in the upper Hubard band}
\label{maa}
\end{center}
\end{figure}

Keeping in mind the scope of the theories used in our approach, we will now investigate
the influence of the additional $p$-band.  In reference \cite{PHW98} these theories are
thoroughly evaluated. To gain the best possible
comparison with these calculations, we choose the same lattice structure
(fcc-tight-binding, $d\rightarrow\infty$, after particle-hole transformation)
for our investigations of the two-band model. The density of states
reads
\begin{equation}
\rho^{(0)}(E)=\frac{e^{-(1+\sqrt{2}E/t^*)/2}}{t^*\sqrt{\pi(1+\sqrt(2)E/t^*)}}\quad
.
\label{DOS}
\end{equation}
In the following all
energies will be given in units of $t^*$.The density of states exhibits a divergence at the lower
band edge. This feature is known to stabilizes
ferromagnetism. The main trends regarding the
influence of the hybridization are also present in
other lattice structures (e.g. sc or  bcc tight-binding).
Ferromagnetism, however, is most certain within
the fcc-lattice\cite{WBS98}. 
Fig. \ref{para} and
Fig. \ref{maa} show the
quasiparticle densities of states calculated with {\it SDA} and 
{\it MAA} for different values of $V$ in the paramagnetic
case. In both theories the
QDOS consists of two Hubbard-bands and the uncorrelated band. These
bands move apart with rising $V$ while the
correlated sub-bands are broadened. One can see that the band shifts are
proportional to $V^2$, which agrees perfectly with the results
for free bands (\ref{dispersionen2}, \ref{shift}, \ref{broad.}) and the two-site cluster (\ref{cluster}).\\

\begin{figure}
\begin{center}
      \epsfig{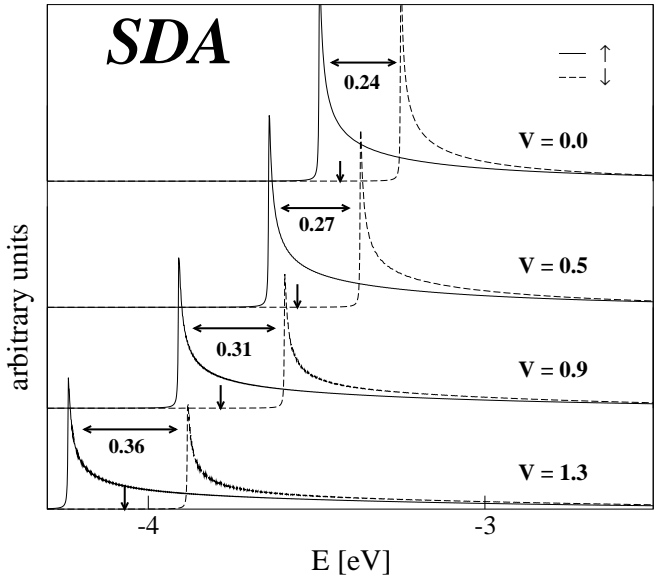}
\caption{Lower Hubbard band of the $\uparrow$- and $\downarrow$-
  correlated QDOS in
  the ferromagnetic phase. Parameters as in Fig. \ref{para} and Fig. \ref{maa}.
  The distance between the $\uparrow-$ and $\downarrow-$band increases
  with $V$.} 
\label{ferro}
\end{center}
\end{figure}

\begin{figure*}
\begin{center}
      \epsfig{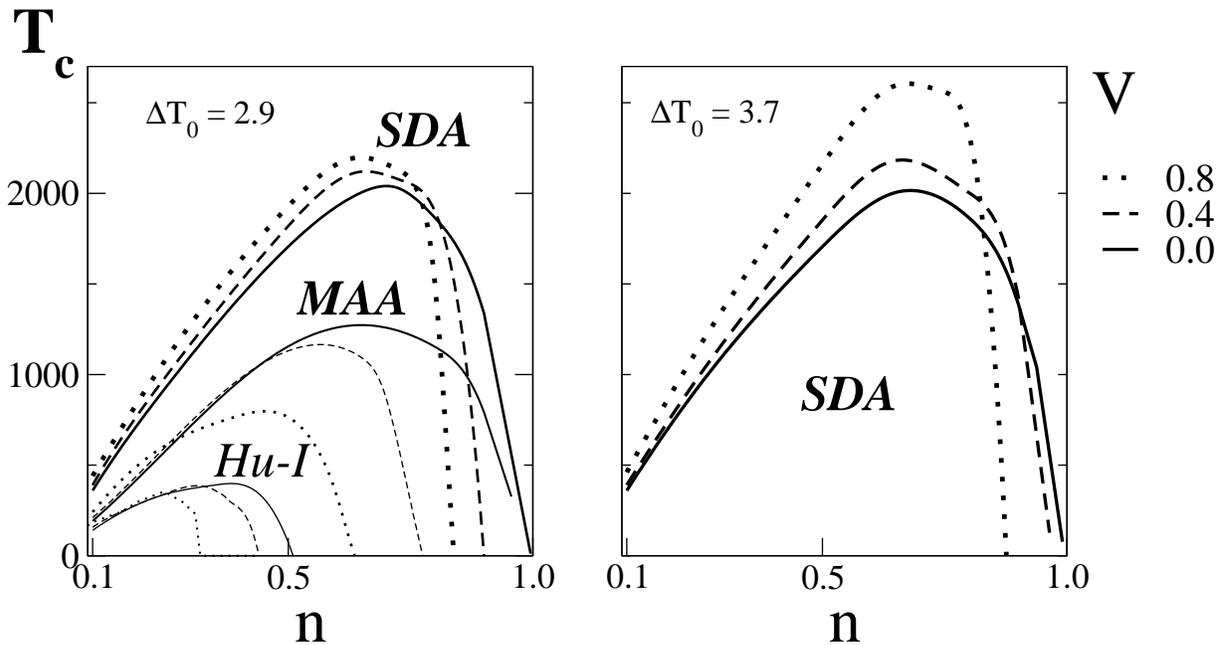}
\caption{n-T phase diagram for different Values of $V$ and different
  band distances $\Delta T_0$. Other parameters as in the previous figures.}
\label{tvonn}
\end{center}
\end{figure*}

\begin{figure}
\begin{center}
      \epsfig{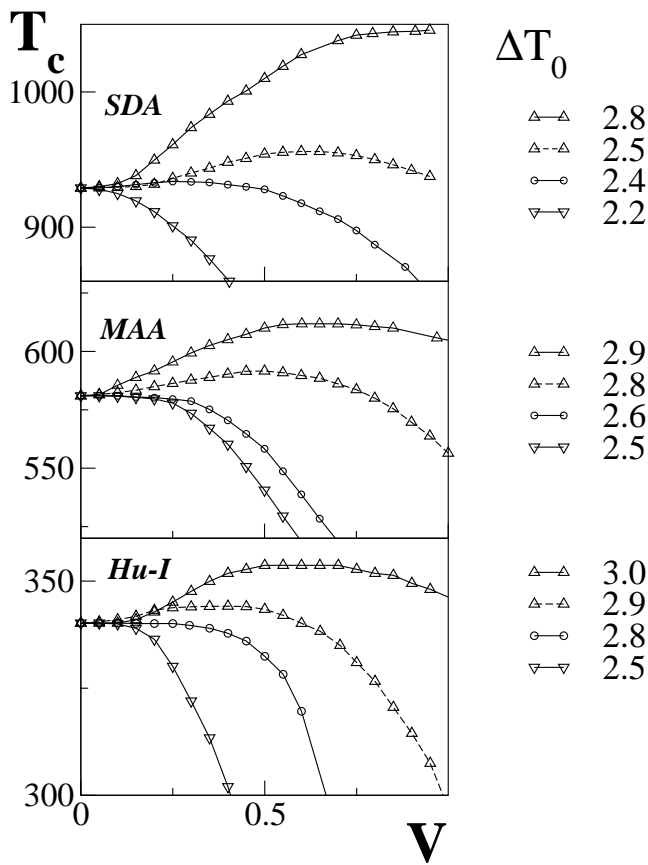}
\caption{Curie temperatures in dependence of $V$ calculated within
  different theories. Triangles up: stabilization of ferromagnetism for
  small $V$, triangles down: destabilization of ferromagnetism. Other
  parameters as in the previous figures.}
\label{tvonV} 
\end{center}
\end{figure}

\begin{figure}
\begin{center}
      \epsfig{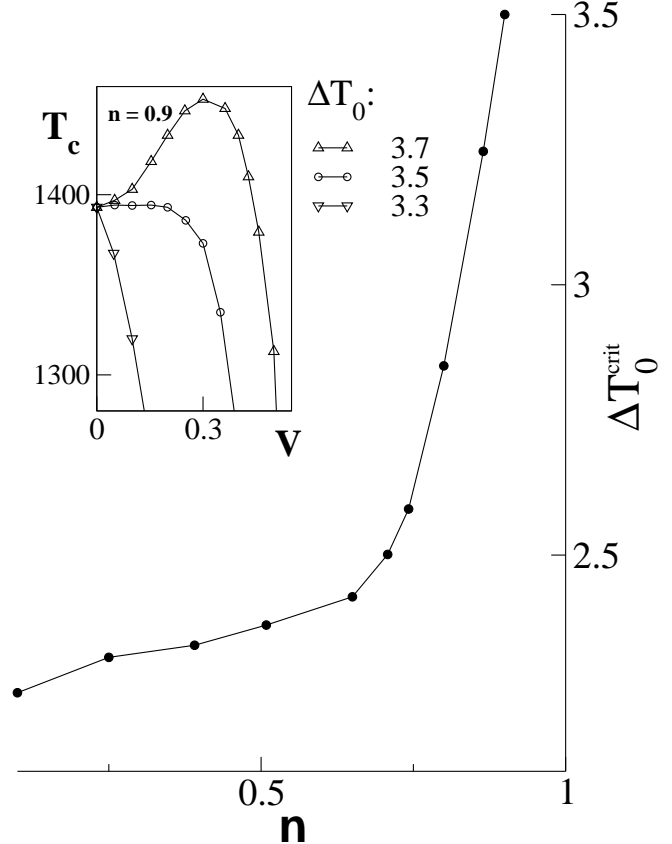}
\caption{Critical band distance that separates the stabilizing and the
  destabilizing regime. Above the line ferromagnetism is stabilized by
  the uncorrelated band for small $V$. Below the line ferromagnetism is
  destabilized. Inset: as in Fig. \ref{tvonV} for $n=0.9$,
  {\it SDA}. Further parameters as in the previous figures.}
\label{efcrit}
\end{center}
\end{figure}

\begin{figure}
 \begin{center}
    \epsfig{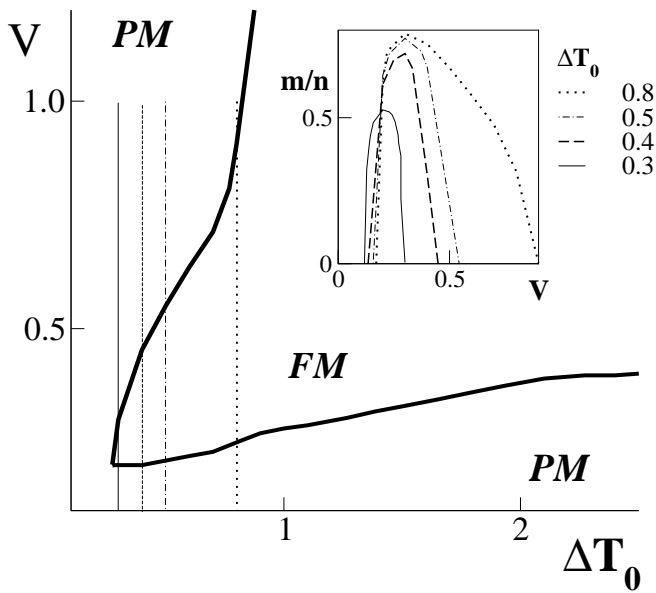}
\caption{$p$-band induced ferromagnetism for a
  bcc tight-binding lattice. Inset: Magnetization in dependence of $V$.
  Parameters: $T=0K,\, U=4,\,{\cal W}_0=1,\,\alpha=4,\,n=0.55,\,(SDA).$}
\label{ef_V}
    \end{center}
\end{figure}

Fig. \ref{ferro} displays the lower Hubbard band ({\it SDA}) of the correlated
density of states in the ferromagnetic case.
 It turns out, that the hybridization-caused band shift
is in fact spin-dependent in the ferromagnetic phase. The shift is
larger for majority-spin electrons. The
magnetic properties can thus be changed drastically due to the presence of
the uncorrelated band. Within the framework of the {\it SDA} we can derive analytical expressions for
the shift and the broadening by calculating the poles of (\ref{matrix}) with
$\Sigma_{k\sigma}=\Sigma_\sigma^{SDA}$. For $U\rightarrow\infty$ we
find:
\begin{eqnarray*}
E_{1k\sigma}^{SDA}&=&T_0^d+U+(\epsilon_k^d-T_0^d)n_{-\sigma}^d+B_{-\sigma}^{2b}(1-n_{-\sigma}^d)\nonumber\\
E_{2k\sigma}^{SDA}&=&\epsilon_k^p+ V^2 X_{k\sigma}\nonumber\\
E_{3k\sigma}^{SDA}&=&T_0^d+(\epsilon_k-T_0^d)(1-n_{-\sigma}^d)+B_{-\sigma}^{2b}n_{-\sigma}^d-
V^2 X_{k\sigma}
\end{eqnarray*}
where 
\begin{equation*}
V^2
X_{k\sigma}=V^2\frac{1-n_{-\sigma}^d}{{\epsilon}_k^p-T_0^d-(\epsilon_k^d-T_0^d)(1-n_{-\sigma}^d)-B_{-\sigma}^{2b}n_{-\sigma}^d}
\end{equation*}
describes the whole influence of the hybridization. The other terms
are the well-known {\it SDA}-result for the single-band Hubbard model.
For band fillings smaller than unity, the most important energies are $E_{3k\sigma}$ forming the lower Hubbard-band. For non-overlapping
bands (i.e. $\Delta T_0> {\rm max}(\epsilon_k^p-T_0^p)$), we can
rewrite $E_{3k\sigma}^{SDA}$ in terms of a band shift $\Delta T_{V\sigma}$ and a
band broadening factor $x_{V\sigma}$
\begin{eqnarray*}
E_{3k\sigma}^{SDA}&=&T_0^d+\Delta
T_{V\sigma}+n_{-\sigma}^dB_{-\sigma}^{2b}\\
&{}&+(\epsilon_k^d-T_0^d)(1-n_{-\sigma}^d)\cdot
x_{V\sigma}\quad.
\end{eqnarray*}
Both are spin dependent:
\begin{eqnarray}
\Delta T_{V\sigma}&=-&\frac{V^2}{\Delta T_0-n_{-\sigma}^dB_{-\sigma}^{2b}}(1-n_{-\sigma}^d)\nonumber\\
x_{V\sigma}&=&1+\frac{V^2}{(\Delta T_0-n_{-\sigma}^d
  B_{-\sigma}^{2b})^2}(\alpha-1+n_{-\sigma}^d).
\label{effekte}
\end{eqnarray}
$\alpha$ is the ratio of the free band widths as defined in (\ref{definition}).
Thus an hybridization with an uncorrelated band causes alterations in
the band structure similar to the non-interacting case (\ref{shift},
\ref{broad.}), i.e. a band shift and a band broadening for the correlated
sub-band. The important
difference is the {\it spin dependence} of both quantities in the full
system. Equation (\ref{effekte}) describes two competing
effects. The shift to lower energies $\Delta T_{V\sigma}^d$ supports magnetism since it is larger for majority-spin electrons.
The band broadening $x_{V\sigma}$, in contrast, destabilizes magnetism, since
broader bands are known to be inconvenient for band-ferromagnetism. In addition, the spin
dependence of $x_{V\sigma}$ works against ferromagnetism.\\
$\Delta T_{V\sigma}$ and $x_{V\sigma}$ constitute the main mechanisms by
which the $p$-band influences the $d$-band magnetism.\\
In Fig. \ref{tvonn} Curie temperatures are shown in dependence of the
band filling $n^d$ for different parameters V.
 We find
both, stabilization for lower particle densities as well as
destabilization for higher ones in all
theories. Surprisingly, the stabilization is clearly more pronounced, if the band distance
increases (r.h.s of Fig. \ref{tvonn}).\\
Fig. \ref{tvonV}
gives a systematic overview of the $V$-dependence of the Curie
temperature for different band distances $\Delta T_0$.
There exists a critical band
distance $\Delta T_0^c$ that separates regimes with qualitatively different
behaviour of the Curie temperature (lines with circles). This distance
is  about $2.8\,eV$ for {\it Hu-I}, for {\it MAA} approximately $2.6\,eV$, while within the {\it SDA} the critical band
distance is somewhat smaller than $2.4\,eV$. $\Delta T_0^c$ is
characterized by the following:\\
(i) 
Above the critical band distance (triangles up) we are in the
stabilizing regime. Here ferromagnetism can be stabilized by the
uncorrelated bands for small hybridizations
$V$. The Curie temperature $T_c$ shows a maximum as a function of
the hybridization $V$.\\ 
(ii)
The situation is different for small band distances $\Delta T_0<\Delta
T_0^c$ (triangles down). We are now in the destabilizing regime. No
enhancement of the Curie temperature is found,
only destabilization of magnetism.\\
The different behaviour can be understood by
inspecting again the two competing band structure effects (\ref{effekte}) of
the hybridization: At small band distances $\Delta T_0$ the
destabilizing band broadening $x_{V\sigma}\sim 1+\frac{V^2}{\Delta T_0^2}$
is more important than the stabilizing shift $\Delta
T_{V\sigma}\sim-\frac{V^2}{\Delta T_0}$. We are in the destabilizing
regime. As $\Delta T_0$ increases, the
shift more and more over-compensates the broadening, though both effects
become smaller. Thus ferromagnetism can be stabilized for $\Delta
T_0>\Delta T_0^c$. Large inter-band fluctuations at high values of $V$, however,
suppress ferromagnetism also in this regime.\\
For further investigations we only show 
{\it SDA}-results since the
dependence of the hybridization strength $V$ is qualitatively the same in all theories. First we
want to look at the critical band distance, that separates the
stabilizing from the destabilizing regime. It depends
sensitively on the band filling $n^d$ (Fig. \ref{efcrit}). 
For densities
$n^d$ closer to half filling the
  critical band distance is enhanced. This
  could reflect the fact, that the Fermi energy rises with increasing band
  filling and therefore the gap between the Fermi energy and the
  uncorrelated states becomes smaller. This enhances the inter-band fluctuation rate and the stabilization of ferromagnetism is more unlikely. As in the
  single-band model, no ferromagnetism was found at $n^d\ge 1$ for the free
  density of states (\ref{DOS}).\\
Finally we want to study ground state properties:
The $p$-band can induce a ferromagnetic ground-state, if the single-band
system is paramagnetic, but close to a ferromagnetic transition (Fig. \ref{ef_V},
bcc tight-binding lattice). A ferromagnetic ground
state is induced by the $p$-$d$ hybridization for band distances greater
than $\Delta T_0^c=0.3\,eV$ and for moderate values of $V$. The
magnetization shows a distinct maximum as a function of the
hybridization strength as shown in the inset of
Fig. \ref{ef_V}.\\

\section{Conclusions}
\label{conclusions}
Let us summarize our findings.
The question, whether and how an uncorrelated band can stabilize band
ferromagnetism can now be answered:\\
Stabilization of ferromagnetism is only found for small hybridization
strengths. Strong fluctuations between the bands generally suppress
ferromagnetic order. Small fluctuations can
stabilize ferromagnetism if the band distance is
larger than a critical energy $\Delta T_0^c$, which depends sensitively on the band filling $n$ and on the shape of the free density of
  states.\\
The stabilization and the
  destabilization result from spin-dependent inter-band particle
  fluctuations. They induce a spin-dependent renormalization of the correlated
  quasiparticle density of states. This
  renormalization can be analyzed in terms of a band
  broadening dominating at small band distances and a band shift dominating at larger ones. The former turns out to suppress, the
  latter to stabilize ferromagnetism.\\ In other words: as usual the
  system lowers its energy by inter-band particle fluctuations. Because
  the latters can be spin-dependent, the energy-win is
  different for the spin-up and the spin-down electrons. This in turn
  influences the stability of the ferromagnetic phase and e.g. the
  Curie-temperatures. The described mechanism can also give a "final kick" to a
  system, that is close to a ferromagnetic transition.\\
There are various arguments that compared to this mechanism indirect
exchange interactions as an
RKKY-like coupling of localized $d$-moments are of minor importance:\\
(i) In most of the calculations shown here, the lower band edge
  of the $p$-band is located above the Fermi-energy (Fig. 1 -
  Fig. 6). Except for the mixing of $d-$ and $p-$states, this band is
  therefore empty and no polarization of the $p$-band can be
  expected. This excludes RKKY-coupling.\\
(ii) RKKY-coupling results from simultaneous fluctuations at
  different sites and is thus of order $V^4$ (see e.g. reference \cite{TJF97}).
The same holds for other indirect exchange mechanisms as e.g. super-exchange. 
On the other hand the band-structure effects we discussed above are based on
 uncoupled fluctuations and are thus of order $V^2$. Therefore they will
 dominate the system.

In conclusion we have found a considerable influence of the $p$-band to
$d$-band ferromagnetism in our model.\\
The involved processes are due to the interplay of correlation and hybridization.
Hence our investigations showed that the
$p$-$d$ hybridization should be taken into account in model
calculations to achieve a
realistic description of real substances. Otherwise magnetic properties
may be over- or underestimated.\\ If and how the $p$-bands
influence the anti-ferromagnetic phase was left open in our analysis
and remains an interesting question to be answered in further investigations.
\\\\
This work was supported by the Deutsche Forschungsgemeinschaft within
the Sonderforschungsbereich 290 ("Metallische d\"unne Filme: Struktur,
Magnetismus und elektronische Eigenschaften")


\end{document}